\documentstyle[aps,preprint,tighten]{revtex}

\begin{document}

\draft

\title{Bound states of a spin-$1/2$ charged particle in a magnetic flux tube}
\author{R.~M.~Cavalcanti\footnote{E-mail: rmc@itp.ucsb.edu}$^{\ddag}$
and C.~A.~A.~de Carvalho\footnote{E-mail: aragao@if.ufrj.br}$^{\S}$}
\address{$^{\ddag}$Institute for Theoretical Physics, University of California, 
Santa Barbara, CA 93106-4030, USA \\
$^{\S}$Instituto de F\'\i sica, Universidade Federal do Rio de Janeiro, 
Caixa Postal 68528, Rio de Janeiro, RJ 21945-970, Brasil}
\maketitle

\begin{abstract}

We study a spin-$1/2$ charged particle with gyromagnetic factor
$g>2$ moving in a plane threaded by a magnetic flux tube. We prove
that, if the magnetic field (i) has radial symmetry, 
(ii) has compact support and (iii) does not change sign, there is
at least one bound state for each angular momentum $\ell$
in the interval $[0,v]$, where $v$ (assumed positive) is the total
magnetic flux in units of the quantum of flux.

\end{abstract}




Let us consider a spin-$1/2$ particle with charge $e$ and
gyromagnetic factor $g$, moving in a plane threaded by a magnetic
field. Its dynamics is governed by the Pauli Hamiltonian, 
which, when specialized to the case where the magnetic
field points in the $z$ direction, can be written as 
[$2m=\hbar=c=1$; $\sigma_z={\rm diag}(1,-1)$]
\begin{equation}
H=(-i\nabla-e{\bf A})^2-\frac{1}{2}\,geB\sigma_z
\equiv{\rm diag}(H_{\uparrow},H_{\downarrow}).
\end{equation}

For $g=2$ --- the value which emerges naturally when one derives
the Pauli Hamiltonian as the non-relativistic limit of the
Dirac Hamiltonian ---, it is possible to show\cite{Aharonov,Jackiw,Musto} 
that the above Hamiltonian has zero-energy 
normalizable\footnote{Such states are not normalizable if the magnetic
field has a $\delta$-function singularity\cite{Moroz,Cavalcanti}.}
eigenstates if the total magnetic flux
$\Phi\equiv\int B({\bf x})\,d^2x$ is larger than $\Phi_0=2\pi/|e|$
(in general, there are $N$ such states if $N\Phi_0<\Phi\le (N+1)\Phi_0$).

On the other hand, the experimental value of the electron $g$-factor is
not 2, but $2.0023$ (this ``anomaly'' can be understood as a result
of QED radiative corrections\cite{Schwinger}). This has motivated
some investigations of the case $g>2$ in the last few years. In this case,
qualitatively new phenomena occur: (a) an arbitrarily small (but nonzero) 
magnetic flux $\Phi$ can create a negative-energy 
state\cite{Bordag,Moroz,Cavalcanti}, and (b) even if $\Phi=0$,
bound states can appear if the magnetic field is strong 
enough\cite{Bentosela}. 

A very important limitation of (a) is that it has been proved only
for a few types of magnetic field profiles, and so it is not known
to what extent it is a general result.
The purpose of this paper is to partially fill this gap. We shall
prove that, if $g>2$ and the magnetic field satisfies the following
conditions:
(i) it has radial symmetry, 
(ii) it has compact support (and so the flux tube has a well defined radius R)
and (iii) it does not change sign (without loss of generality, we 
shall suppose that $eB(r)\ge 0$), 
then $H_{\uparrow}$ has at least $[v]+1$ eigenstates 
with negative energy, where $v\equiv\Phi/\Phi_0$ and $[v]$ denotes its integer
part (i.e., $v=[v]+\epsilon$, $0\le\epsilon<1$).\footnote{Here, and in what 
follows, we are assuming that $v>0$. 
$v=0$ corresponds to the absence of a magnetic field, in which case 
there are no bound states (instead of $[v]+1$).}
In order to prove this assertion, we first note that, because of (i), 
we can choose the vector potential
in the symmetric gauge, ${\bf A}=A_{\varphi}(r)\,{\bf e}_{\varphi}$, in terms
of which the magnetic field is written as 
\begin{equation}
B(r)=\frac{1}{r}\,\frac{d}{dr}\,[rA_{\varphi}(r)].
\end{equation}
In this gauge, the time-independent Schr\"odinger equation
is separable in polar coordinates $(r,\varphi)$, and 
one can perform a partial-wave decomposition.

Now, let us consider the following variational wave functions:
\begin{equation}
\psi_{\ell}(r,\varphi)={\rm e}^{i\ell\varphi}\times\cases{
	a_{\ell}\,r^{\ell}\,{\rm e}^{-e\phi(r)-\alpha r}& for $0<r<R$, \cr
	b_{\ell}\,K_{|\ell-v|}(\mu r)\quad(\mu>0) & for $r>R$,}
\end{equation}	
where $\phi(r)\equiv\int_0^r A_{\varphi}(r')\,dr'$ and $K_{\nu}(z)$
is the modified Bessel function\cite{Abramowitz}. 
The coefficients $a_{\ell}$ and $b_{\ell}$ are chosen
in such a way that $\psi_{\ell}(r,\varphi)$ is continuous at $r=R$
and $\langle\psi_{\ell}|\psi_{\ell}\rangle=1$.
The latter condition can only be fulfilled if $\ell\ge 0$, in
order that $|\psi_{\ell}|^2$ be integrable at the origin 
(integrability at infinity is ensured by the positivity of $\mu$).
We shall also impose that $\partial_r\,\psi_{\ell}(r,\varphi)$ be
continuous at $r=R$; together with the continuity of
$\psi_{\ell}(r,\varphi)$, this implies that $\mu$ must satisfy the
following equation:
\begin{equation}
\label{match}
\frac{\mu R\,K'_{|\ell-v|}(\mu R)}{K_{|\ell-v|}(\mu R)}
=-v+\ell-\alpha R,
\end{equation}
where we have used the identity
\begin{equation}
\phi'(R)=A_{\varphi}(R)=\frac{1}{R}\,\int_0^R B(r)\,rdr=\frac{v}{eR}.
\end{equation}
Since the l.h.s.~of (\ref{match}) tends to $-|\ell-v|$ when
$\mu\to 0^+$, and it behaves asymptotically as $-\mu R$ when
$\mu\to\infty$\cite{Abramowitz}, that equation will have a
positive root if $\ell\le v$ and $\alpha>0$.

We have thus obtained $[v]+1$ normalizable
functions $\psi_{\ell}(r,\varphi)$ ($\ell=0,1,\ldots,[v]$)
which, together with $\partial_r\,\psi_{\ell}(r,\varphi)$,
are continuous at $r=R$. Now, we shall show that,
for each of these functions,
it is possible to find a positive $\alpha$ such that
$\langle\psi_{\ell}|H_{\uparrow}|\psi_{\ell}\rangle<0$. 
It then follows from the variational principle that there 
exists (at least) one negative energy eigenstate of $H_{\uparrow}$
with angular momentum $\ell$. In fact,
a direct calculation shows that
\begin{eqnarray}
\langle\psi_{\ell}|H_{\uparrow}|\psi_{\ell}\rangle&=&
2\pi|a_{\ell}|^2\int_0^R
\left[\frac{(2\ell+1)\alpha }{r}-2\alpha eA_{\varphi}(r)
-\alpha ^2-\frac{1}{2}\,(g-2)eB(r)\right]
r^{2\ell}{\rm e}^{-2e\phi(r)-2\alpha r}\,rdr
\nonumber \\
& &-2\pi|b_{\ell}|^2\mu^2\int_R^{\infty}K_{|\ell-v|}^2(\mu r)\,rdr
\nonumber \\
&\le&2\pi\,|a_{\ell}|^2\int_0^R
\left[\frac{(2\ell+1)\alpha }{r}-\frac{1}{2}\,(g-2)eB(r)\right]
r^{2\ell}{\rm e}^{-2e\phi(r)-2\alpha r}\,rdr.
\end{eqnarray}
Our goal will be accomplished if we choose a positive $\alpha$
satisfying
\begin{equation} 
\label{alpha}
\alpha <\frac{(g-2)e\int_0^R B(r)\,r^{2\ell+1}
{\rm e}^{-2e\phi(r)-2\alpha r}\,dr}
{2(2\ell+1)\int_0^R r^{2\ell}{\rm e}^{-2e\phi(r)-2\alpha r}\,dr}.
\end{equation}
That such a choice is possible follows from the fact that the
r.h.s.~of the above inequality is a continuous function of $\alpha$,
and (if $v>0$) it tends to a positive number when $\alpha\to 0$. 
This completes the proof. 

Finally, we make a couple of remarks: (i) an ``extra'' bound state 
(with $\ell>v$) can show up, provided $\ell-v<\alpha R/2$ (in this case too
one can prove that Eq.~(\ref{match}) has a positive root); (ii) our proof
breaks down if $B$ has a $\delta$-function singularity at the
origin; this case requires a special treatment (see discussion in
Refs.\cite{Bordag,Moroz}), and will not be considered here.


\acknowledgments

We thank Eduardo Souza Fraga for useful conversations.
This work was finantially supported by the Conselho
Nacional de Desenvolvimento Cient\'\i fico e Tecnol\'ogico (CNPq)
and, in part, by the National Science Foundation under Grant
No.~PHY94-07194.



\begin{references}

\bibitem{Aharonov} Aharonov Y and Casher A 1979 {\it Phys. Rev.} A {\bf 19}
2461--2

\bibitem{Jackiw} Jackiw R 1984 {\it Phys. Rev.} D {\bf 29} 2375--7

\bibitem{Musto} Musto R, O'Raifeartaigh L and Wipf A 1986 {\it Phys. Lett.} B 
{\bf 175} 433--8

\bibitem{Schwinger} Schwinger J 1948 {\it Phys. Rev.} {\bf 73} 416;
1949 {\it Phys. Rev.} {\bf 76} 790--817; 
1951 {\it Phys. Rev.} {\bf 82} 664--79 

\bibitem{Bordag} Bordag M and Voropaev S 1993 {\it J. Phys. A: Math. Gen.} 
{\bf 26} 7637--49

\bibitem{Moroz} Moroz A 1996 {\it Phys. Rev.} A {\bf 53} 669--94

\bibitem{Cavalcanti} Cavalcanti R M, Fraga E S and de Carvalho C A A 1997
{\it Phys. Rev.} B {\bf 56} 9243--6 

\bibitem{Bentosela} Bentosela F, Exner P and Zagrebnov V A 1998
{\it J. Phys. A: Math. Gen.} {\bf 31} L305--11 

\bibitem{Abramowitz} Abramowitz M and Stegun I A (ed) 1965
{\it Handbook of Mathematical Functions\/} (New York: Dover)

\end{references}
\end{document}